\documentclass[]{article}
\pdfoutput=1 % if your are submitting a pdflatex (i.e. if you have
\usepackage{jheppub}
\usepackage[T1]{fontenc} % if needed
\usepackage{color}
\usepackage{mathtools}
\usepackage[caption = false]{subfig}
\usepackage[normalem]{ulem}

\preprint{UTTG--23--26}

\title{A defect in holographic interpretations
\mbox{of tensor networks}}

\author[a]{Bart{\l}omiej Czech,}
\author[b]{Phuc H. Nguyen}
\author[b]{and Sivaramakrishnan Swaminathan}

\affiliation[a]{Institute for Advanced Study, Princeton, NJ 08540, USA}
\affiliation[b]{Theory Group, Department of Physics and Texas Cosmology Center, The University of Texas at Austin, Austin, TX 78712, USA}

\emailAdd{czech@ias.edu}
\emailAdd{phn229@physics.utexas.edu}
\emailAdd{siva@physics.utexas.edu}

\abstract{We initiate the study of how tensor networks reproduce properties of static holographic space-times, which are not locally pure anti-de Sitter. We consider geometries that are holographically dual to ground states of defect, interface and boundary CFTs and compare them to the structure of the requisite MERA networks predicted by the theory of minimal updates. When the CFT is deformed, certain tensors require updating. On the other hand, even identical tensors can contribute differently to estimates of entanglement entropies. We interpret these facts holographically by associating tensor updates to turning on non-normalizable modes in the bulk. In passing, we also clarify and complement existing arguments in support of the theory of minimal updates, propose a novel ansatz called rayed MERA that applies to a class of generalized interface CFTs, and analyze the kinematic spaces of the thin wall and AdS$_3$-Janus geometries.
}

\begin{document}
\maketitle
%\flushbottom
%\newpage

\section{Introduction}
\label{sec:intro}

In the last decade, two of the most successful approaches to studying conformal field theories---holographic duality and tensor networks---have turned out to be intimately tied to entanglement. In the AdS/CFT correspondence \cite{adscft, adscftwitten}, the Ryu-Takayanagi proposal \citep{rt,rt2} revealed that holographic spacetimes function as maps of CFT entanglement; meanwhile, the Multi-scale Entanglement Renormalization Ansatz (MERA) \citep{mera,mera2} arose largely from considering the scale dependence of entanglement entropies in conformal field theories. The fact that quantum entanglement plays a clarifying role in both approaches suggests that holographic spacetimes and MERA networks may be linked by a more direct relationship.

Such a relationship was first proposed by Swingle~\cite{swingle, brian2012} (see also \cite{ehm, cmeraads, ehm2, cmeraads2, lqg2ehm}) who pointed out that the MERA network for a CFT ground state bears a striking resemblance to the geometry of anti-de Sitter (AdS) space. An alternative proposal~\cite{kinematicsp, secondpaper} argued that the translation between MERA and holography is mediated by an auxiliary construct termed kinematic space. But both proposals are largely qualitative and would benefit from a broader class of examples, other than the case of the CFT vacuum / pure AdS geometry.
Some steps in that direction were taken in refs.~\cite{secondpaper, BTZpaper} (see also refs.~\cite{swingle, brian2012, donm}) who compared MERA representations of CFT$_2$ thermal states to the BTZ geometries. The analysis in ref.~\cite{secondpaper} also included Virasoro descendants of the CFT vacuum, related to other locally AdS$_3$ space-times. One commonality of all these examples is that they rely on the extended conformal symmetry in two dimensions. To further explore how MERA and the holographic duality may come together, we need to consider holographic duals and MERA representations of CFT states, which are not related to the vacuum by the application of an anomalous symmetry.

This is the subject of the present paper.\footnote{Other tensor network realizations of broader classes of geometries, mostly set in the context of the ER=EPR \cite{erepr} and the complexity=action \cite{complexity} conjectures, include \cite{hartmanmaldacena, shocks, notenough, chaoschannels, complexityaction, Chua:2016cxo, May:2016dgv, Leutheusser:2016awv}. Those works concentrate on the dynamics of space-times while our interest here is on bulk duals of ground states of more general classes of CFTs.} We consider the ground states of two-dimensional conformal field theories whose global symmetry has been broken from $SO(2,2)$ down to $SO(2,1)$ by the presence of a defect, an interface or a boundary.\footnote{In order not to clutter the text, we will refer to all these setups as `defects' unless the context requires distinguishing defects, interfaces and boundaries.} On the tensor network side, the `theory of minimal updates'~\cite{minupdates} governs the structure of the MERA representations of such states. In holography, there have been many discussions and several explicit examples of holographic defect/interface~\cite{Karch:2000gx, locallylocalized, Bak:2007jm} and boundary CFTs~\cite{Takayanagi:2011zk, Fujita:2011fp, ugajin} in two dimensions. Our goal is to compare these MERA networks and holographic geometries and analyze in what way, if at all, they relate to one another.

Our principal findings are the following:
\begin{enumerate}
\item In Sec.~\ref{sec:minupdates}, we complement existing arguments~\cite{impurityalgorithms, minupdates} which support the validity of the minimally updated MERA and clarify the circumstances under which it is expected to hold. It applies to actual defect and interface CFTs, but not to generic two-dimensional theories with $SO(2,1)$ symmetry.
\item In Sec.~\ref{sec:rayedMERA}, we propose {\bf rayed MERA}---a simple generalization of MERA, which should capture ground states of generic two-dimensional theories with $SO(2,1)$ symmetry. In holography, the cases where the minimally updated MERA suffices versus those requiring rayed MERA are distinguished by the boundary region where non-normalizable modes are supported.
\item In Sec.~\ref{sec:holography}, we discuss two examples of holographic defect CFTs. We conclude that a na{\"\i}vely local relation between MERA networks and AdS$_3$ geometries, in which a specific region of the MERA network corresponds to a specific region of (the spatial slice of) AdS$_3$, does not hold. This applies both to the direct AdS-MERA correspondence of refs.~\cite{swingle, brian2012} and to the kinematic proposal of refs.~\cite{kinematicsp, secondpaper}.
\item Instead, a key ingredient in relating tensor networks to holographic geometries is that every bond should be associated with the amount of entanglement across it and not with more na{\"\i}ve measures such as the bond dimension. This point was already made in ref.~\cite{secondpaper}; here we exemplify it.
We expect this conclusion to apply to all tensor network models of holography, not just to MERA.\@
\end{enumerate}
Combining these observations leads to the following holographic interpretation of the prescription of \cite{minupdates}: the theory of minimal updates specifies which tensors do / do not register the effect of turning on non-normalizable modes in the bulk. We expand on this statement and put our work in a broader context in the Discussion section.

%Combining these observations leads to the following statement: in a holographic interpretation of a ground state MERA, non-normalizable modes change the tensors of the network while normalizable modes keep the tensors unaltered. We expand on this statement and put our work in a broader context in the Discussion section.

\paragraph{Remarks:} Our paper assumes a familiarity with MERA, the AdS/CFT correspondence and their mutual connections. A good review of MERA is ref.~\cite{mera-review}, reviews of AdS/CFT include ref.~\cite{5auth, Nastase:2007kj, Ramallo:2013bua} while relevant discussions of parallels between MERA and the holographic duality include refs.~\cite{swingle, guifrehologr, secondpaper} (see also \cite{ningcritique}). MERA is a variational ansatz and gives the description of a desired state only after optimization; throughout the text we will be referring to optimized MERA networks. We also assume that all the gauge freedom in the network was used to exhibit it in a maximally symmetric form.

\section{Minimal Updates}
\label{sec:minupdates}

Consider a 1+1-dimensional CFT deformed by a localized defect. The defect traces a 0+1-dimensional world-line and introduces a preferred location in space. Thus, it breaks the global symmetry from $SO(2,2)$ down to $SO(2,1)$ or a subgroup thereof. We are interested in theories, where the full $SO(2,1)$ consistent with a defect is preserved. We shall refer to such theories as dCFTs, though it should be remembered that this class of theories includes interface and boundary CFTs. We emphasize that the symmetries of dCFTs do not include translations (broken by the defect), but do include scale transformations centered at points on the defect world-line.

\begin{figure*}[t]
\begin{center}
    \includegraphics[width=12cm]{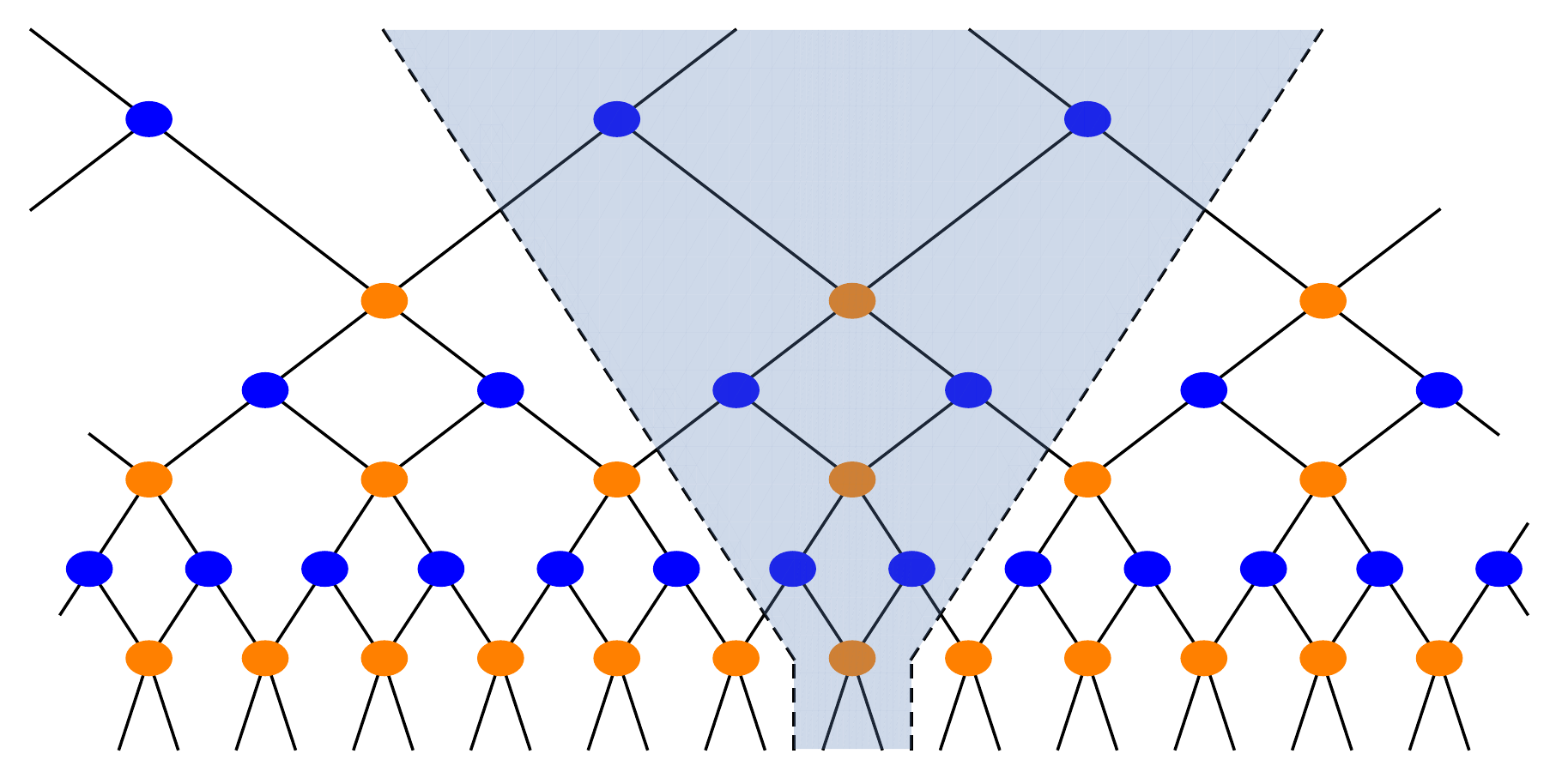}
\end{center}
\caption{The minimal updates prescription (MUP): When we deform a CFT by a defect, only the tensors in the `causal cone' of the defect (shaded blue) need to be replaced in order to account for the defect.}
\label{fig:MinimalUpdateMERA}
\end{figure*}

The minimal updates proposal (MUP)~\cite{minupdates} is a simple tensor network ansatz for the ground state wavefunction of a dCFT.\@ As an input, it starts with an optimized MERA network representing the ground state of the undeformed (parent) CFT$_2$. The MUP asserts that a dCFT ground state can be captured
by `updating' in the input MERA only those tensors, which live in the causal cone\footnote{The `causal structure' in MERA was introduced in ref.~\cite{mera2}; see ref.~\cite{secondpaper} for a discussion relevant to holographic duality.} of the defect location (see Fig.~\ref{fig:MinimalUpdateMERA}).

The MUP is a remarkably powerful ansatz. The computational simplifications owed to reusing the undeformed CFT ground state MERA are enormous. Empirically, the MUP achieves a remarkable accuracy on benchmark examples~\cite{impurityalgorithms,minupdates}, including the case of topological defects~\cite{hauru}.

\subsection{Rationales for Minimal Updates}

Two rationales have been offered by its authors in support of minimal updates.

First, minimal updates guarantee that a local defect remains local after coarse-graining~\cite{minupdates}. As explained in footnote~36 of that reference, initially allowing the update to extend away from the causal cone will, after optimization, lead to a generally location-dependent set of tensors: a defect not confined to a causal cone can `spill out' under renormalization. The motivation behind MUP is to forestall this undesirable scenario. This rationale, however, is not a proof of validity. The symmetry of the problem does not guarantee that tensors in the description of a dCFT ground state are location-independent (see Sec.~\ref{sec:rayedMERA} below.)

Second, there is an algorithmic procedure which takes a discretized (Trotter-Suzuki) version of the Euclidean path integral and transforms it into a MERA representation of the ground state~\cite{tnr}. In effect, Tensor Network Renormalization (TNR) is a derivation of MERA.\@ Applied to Euclidean path integrals of dCFTs, TNR can return a MERA network with a structure predicted by MUP~\cite{tnr2mera}. This seems to provide a derivation of minimal updates, but here too there is a caveat.
A key step in TNR is a local substitution of tensors in the discretized path integral, which is justified by bounding the resulting error (cost function) to a desired tolerance. When the cost function takes into account only the local environment of the tensor to be replaced, the TNR algorithm yields the minimally updated MERA.\@ However, as discussed in Sec.~VIII~(B) of ref.~\cite{hauru}, the TNR algorithm with a global cost function may not produce a MERA with the MUP-dictated structure. Since the conditions under which it suffices to work with a local environment are not known, the status of this second rationale for MUP is also unclear.

In summary, refs.~\cite{minupdates} and \cite{tnr2mera} give two independent rationales for the validity of the minimal updates proposal, neither of which is foolproof. Here we offer a third argument, which relies on symmetry and known properties of dCFTs:

\subsection{Minimal Updates and the Boundary Operator Expansion}

A key new ingredient in a dCFT is the appearance of the Boundary Operator Expansion (BOE) \cite{osborn, McAvity:1993ue}:
\begin{equation}
\mathcal{O}_\eta{(x)} = \sum_{i} \frac{B_{\mathcal{O}_\eta}^{\mathcal{\hat{O}}_{\hat{\eta}_i}}}{{(2y)}^{\eta-\hat{\eta}_{i}}} \hat{O}_{\hat{\eta}_i}{(\textbf{x})}
\label{eq:boe}
\end{equation}
Here we set up coordinates $x = (y, \textbf{x})$ where $y$ is the direction perpendicular to the defect and $\textbf{x}$ are the directions along the defect world-volume. Hats mark operators living on the codimension-1 world-volume of the defect. In the formula above we also assumed that $\hat{O}_{\hat{\eta}_i}$ are scaling operators, i.e. they have well-defined scaling dimensions $\hat{\eta}_{i}$ under dilations centered at the defect location. The BOE allows us to decompose the action of any local operator according to irreducible representations of the residual $SO(1,2)$ symmetry.

In an ordinary CFT all correlation functions can in principle be reduced to kinematic invariants multiplied by products of OPE coefficients, which are the only dynamical data in the theory. In a dCFT, there is an analogous statement: the complete set of dynamical data consists of the BOE coefficients $B_{\mathcal{O}_\eta}^{\mathcal{\hat{O}}_{\hat{\eta}_i}}$ together with the familiar OPE coefficients used for fusing operators away from the defect. For example, one-point functions of local operators in a dCFT are generically non-vanishing and can be read off from fusing the local operators with the defect using the BOE:
\begin{equation}
\langle \mathcal{O}_\eta(x) \rangle = \frac{B_{\mathcal{O}_\eta}^{\hat{1}}}{(2y)^\eta}
\label{eq:1pt}
\end{equation}
Similarly, a correlation function of two local away-from-defect operators $\mathcal{O}_{\eta_1}$ and $\mathcal{O}_{\eta_2}$ can be obtained by first fusing them using the OPE into $\mathcal{O}_\eta$ and then applying eq.~(\ref{eq:1pt}) or, in a different channel, by sequentially fusing $\mathcal{O}_{\eta_1}$ and $\mathcal{O}_{\eta_2}$ with the defect via a double application of the BOE.\@

To verify the validity of the minimal updates proposal, we only need to confirm that the ansatz is powerful enough to correctly encode the away-from-defect OPE and the BOE coefficients.
It is well known that the optimized tensors of the ordinary MERA essentially compute the OPE coefficients of a CFT.\@ This is manifest in the way in which OPE coefficients are extracted from MERA; see e.g. \cite{meraope}. By reusing the undeformed CFT ground state MERA, the minimal updates proposal effectively borrows the undeformed theory's OPE coefficients for fusing away-from-defect local operators. Indeed, a ground state ansatz that departs from the minimally updated MERA would contaminate the fusion rules for operators applied away from the defect.

The above logic implies that the role of the updated region is to encode the remaining dynamical data---the BOE coefficients. Is the ansatz powerful enough to do so? As a computational problem, finding the correct update has the same structure (the same set of inputs and outputs) as the problem of finding the OPE coefficients in the familiar applications of MERA to ordinary CFTs. In both cases, we are looking for tensors that represent a super-operator, which fuses two given sets of operators into one. This argument reduces the question of the validity of the MUP for describing dCFT ground states to the long-settled question of whether the ordinary MERA captures ground states of ordinary CFTs.

This justification for the MUP was not spelled out in \cite{minupdates} or subsequent papers, though similar arguments appeared in \cite{glensslides}. We believe it is important to emphasize the relation between minimal updates and the dCFT technology, especially with a view to the following generalization.

\section{Rayed MERA}
\label{sec:rayedMERA}

Thus far we have considered dCFTs---theories obtained from ordinary CFTs by introducing codimension-1 defects. In general, however, the class of two-dimensional theories with $SO(2,1)$ invariance is much larger. One way to obtain such a theory is by a deformation and (if the deformation is not exactly marginal) an RG flow to a new fixed point. To preserve the symmetry, the sources entering the deformation should have a power-law dependence with $y$, the distance from the world-line fixed by the $SO(2,1)$ symmetry. Still more generally, we can consider a more abstract CFT-like theory in which `OPE coefficients' for fusing $\mathcal{O}_i(x)$ and $\mathcal{O}_j(x')$ have an explicit dependence on
\begin{equation}
\xi = \frac{(x-x')^2}{4yy'}\,,
\label{eq:defxi}
\end{equation}
which is the $SO(2,1)$ invariant built from $x$ and $x'$ discussed e.g. in \cite{osborn}. 

Representing the ground state of a generic, two-dimensional, $SO(2,1)$-invariant theory is outside the scope of the minimal updates proposal. For an arbitrary such theory, there may not exist a CFT whose ground state MERA could be appropriately minimally updated. This is most easily recognized when we consider `OPE coefficients' that depend on $\xi$ from eq.~(\ref{eq:defxi}). We observed previously that in the minimally updated MERA, the region that is directly imported from the parent MERA is responsible for correctly merging away-from-defect operators according to the fusion rules of the parent theory. A theory with $\xi$-dependent `OPE coefficients' does not emulate the fusion rules of any parent theory.

Despite the huge freedom in constructing two-dimensional $SO(2,1)$-invariant theories, it is possible to write down a simple MERA-like ansatz, which ought to capture the ground states of such theories. To do so, note that the tensor network is supposed to represent the wavefunction of the theory at an equal time slice. The only generator of $SO(2,1)$ that acts within a time slice builds dilations about the origin---where the `defect' (the world-line fixed by $SO(2,1)$) and the time slice intersect. The action of the conformal group on the MERA network was studied in \cite{BTZpaper} (see also \cite{guifreconf}). It was found that the orbits of dilations about the origin are tensors, which live on rays emanating from the origin. Thus, the invariance under $SO(2,1)$ dictates that all tensors inhabiting the same ray must be identical, though tensors living on different rays may be distinct. Such an ansatz, which we call {\bf rayed MERA}, is displayed in Fig.~2.

\begin{figure}[t!]
\begin{center}
        \includegraphics[width=10cm]{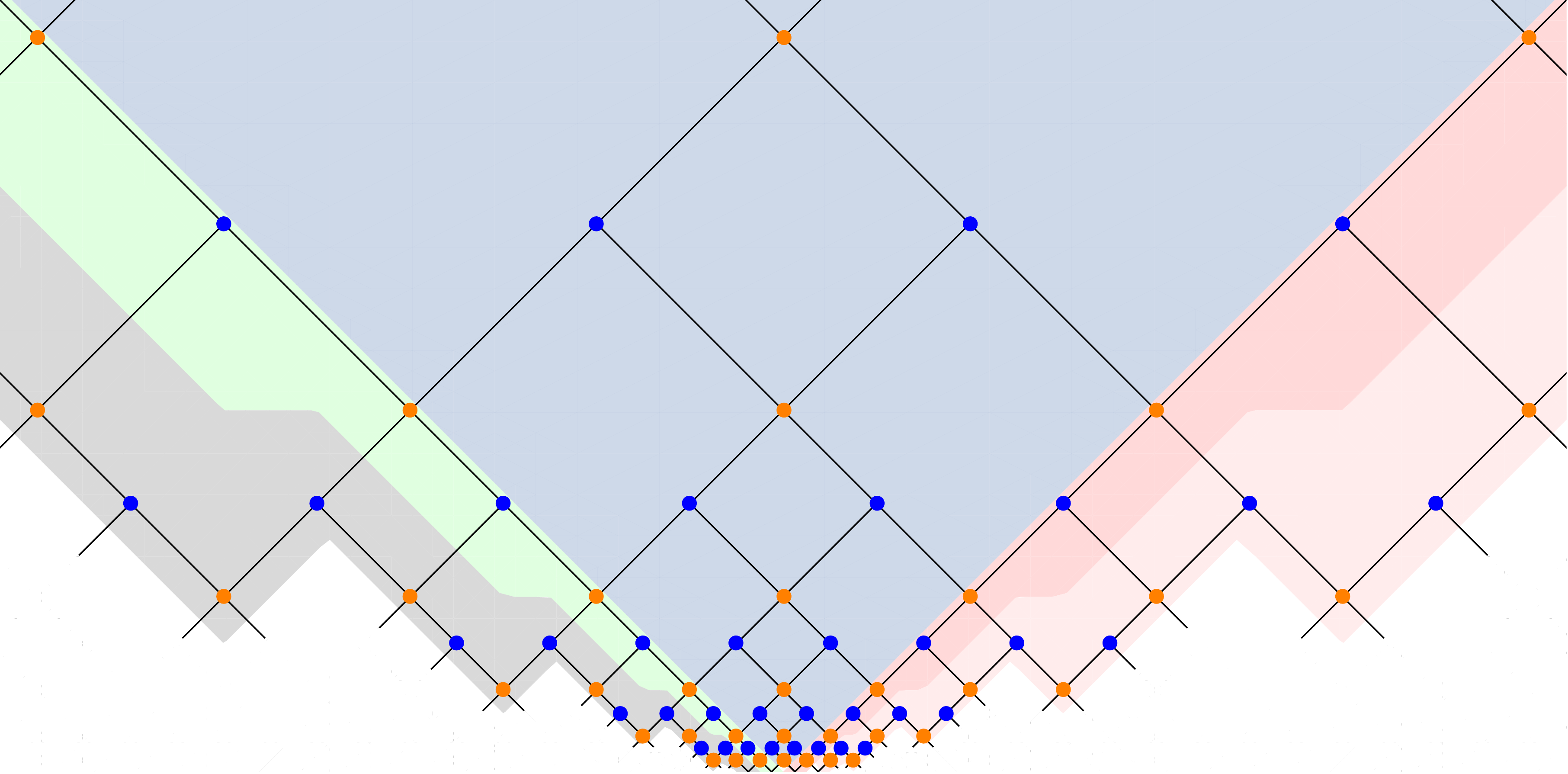}
        \label{RayMERA}
\end{center}
\caption{Rayed MERA: Tensors on each `ray' (color coded) are the same because they are related by a scaling symmetry about the origin (defect location). Tensors inhabiting different rays are in general distinct.}
\end{figure}

Several remarks are in order. First, the minimally updated MERA is a special case of the rayed MERA in which only the vertical ray is distinct from the others. Second, distinct rays are labeled by different values of:
\begin{equation}
\xi = \frac{(x-x')^2}{4yy'}
\quad \xrightarrow[\textrm{(equal time)}]{\textbf{x} = \textbf{x}'} \quad
\frac{(y-y')^2}{4yy'}
\,.
\end{equation}
Here $y$ and $y'$ denote a pair of locations such that if two local operators are inserted there, their causal cones will merge on the ray labeled by $\xi$. If we think of local groups of tensors as encoding OPE coefficients, making the tensors explicitly dependent on $\xi$ amounts to choosing $\xi$-dependent `OPE coefficients.' In the minimally updated MERA, the only $\xi$-dependence distinguishes the parent OPE coefficients from the BOE coefficients, which are encoded on the vertical ray.

\section{Holographic Interpretations}
\label{sec:holography}

We will now look at two holographic realizations of interface CFTs and discuss how, if at all, they relate to either the minimally updated MERA of \cite{minupdates} or the rayed MERA of Sec.~\ref{sec:rayedMERA}. To set the context for our discussion, let us briefly recap how prior proposals related the ordinary MERA to pure anti-de Sitter space.

\paragraph{MERA and holography without defects} Ref.~\cite{swingle} observed a resemblance between the MERA network and a static slice of AdS$_3$, i.e. the hyperbolic disk. Both have a self-similar structure near the cut-off surface and both contain closely related notions of a minimal cut. Geodesics in AdS$_3$, which by the Ryu-Takayanagi proposal compute entanglement entropies of CFT$_2$ regions, resemble minimal cuts through the MERA network. This correspondence is consistent insofar as every bond in a minimal cut through MERA contributes an equal amount to the entanglement entropy of the subtended CFT region. Based on the conclusions of \cite{BTZpaper}, we recognize this fact (first observed in \cite{mera}) as a consequence of the $SO(2,2)$ symmetry of the CFT.

The kinematic proposal of \cite{kinematicsp, secondpaper} instead views individual tensors in MERA as discrete counterparts of geodesics. This does not run into obvious contradictions with \cite{swingle} because every minimal cut in MERA selects a unique tensor, which lives in its top corner. In the kinematic proposal, a key to understanding geodesic lengths and entanglement entropies is the Crofton formula, which schematically reads \cite{kinematicsp}:
\begin{equation}
\textrm{length of a curve} = \int_{\rm intersecting} \mathcal{D}{\rm (geodesics)}.
\label{eq:crofton}
\end{equation}
Here $\mathcal{D}{\rm (geodesics)}$ is the unique measure over the set of geodesics in $\mathbb{H}_2$ invariant under its isometries. The correspondence between MERA tensors and geodesics advocated in \cite{secondpaper} translates eq.~(\ref{eq:crofton}) into simply counting tensors in certain regions of the MERA network.

\subsection{Thin Wall Models: A Na{\"\i}ve Realization of Minimal Updates}

Note that under both holographic interpretations, the directly imported (i.e., not updated) regions of the MUP MERA account for two halves of (the spatial slice of) pure anti-de Sitter space. This is most obvious in the kinematic interpretation: the unaltered regions consist of geodesics with both endpoints on the same side of the defect and both sets (left and right of defect) of such geodesics span one half of the hyperbolic disk. In the original proposal of \cite{swingle}, the minimally updated region should be viewed as a discrete counterpart of a radial geodesic, with one half of $\mathbb{H}_2$ on each side of it. This is because MERA does not accommodate a notion of locality narrower than the width of one causal cone \cite{meraope}. Whichever proposal we adopt, the regions that remain unaltered by the minimal updates should be viewed as two halves of the hyperbolic disk, each ending on a geodesic diameter.

From this observation, one could venture the following, {\bf na{\"\i}ve holographic interpretation} of the minimal updates proposal: that the holographic dual of a dCFT should contain two undeformed halves of pure anti-de Sitter space separated by some `wall.' Whatever the wall is, on either side of it should be (at least) one half of pure anti-de Sitter space.

We shall see later that this holographic reading of the minimal updates proposal is too na{\"\i}ve because it is too stringent. But before that, let us inspect a class of models that realize this na{\"\i}vely stringent interpretation of minimal updates:

\paragraph{Thin wall models} Consider a simple toy model for the holographic dual of a dCFT, which consists of two AdS$_{3}$ patches glued together with a tensionful brane. Such models were discussed for example in \cite{permeable, Chang:2013mca, Jensen:2013lxa, erdmenger} (for early geometric analyses see \cite{bachas, Bachas:2000fr}), building up on an embedding in string theory \cite{Karch:2000gx, locallylocalized}. As we clarify below, the holographic duals of boundary CFTs discussed in \cite{Takayanagi:2011zk, Fujita:2011fp, ugajin} also fall into this class.

\begin{figure}[!htbp]
    \begin{center}
        \includegraphics[width=10cm]{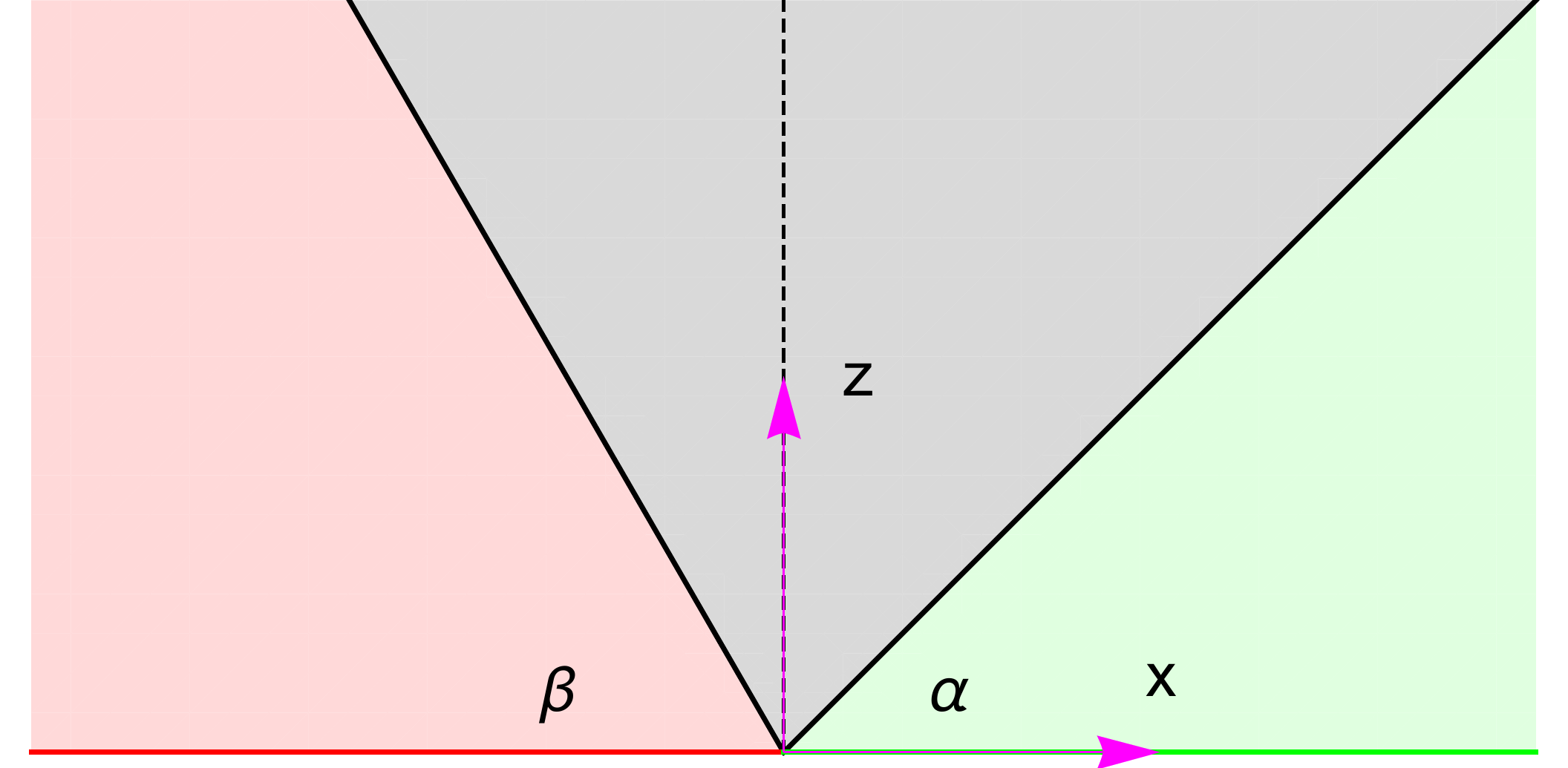}
    \end{center}
	\caption{A thin wall geometry consists of two wedges of pure AdS$_3$ (pink and green regions) glued along a tensionful wall.
The wall occupies a `straight line' in the Poincar{\'e} coordinates, which delimits each AdS$_3$ chunk. The two straight lines are identified.}
\label{fig:gluing}
\end{figure}

The setup is illustrated in Fig.~\ref{fig:gluing}. The two AdS patches can have different curvatures, which would correspond to coupling along an interface two CFTs with central charges $c_L$ and $c_R$. (The special case $c_L = c_R \equiv c$ are actual defect CFTs, as opposed to the more general variety of interface CFTs.) The famous Brown-Henneaux formula \cite{Brown:1986nw} relates the central charges to the radii of curvature:
\begin{equation}
\frac{L}{G} = \frac{2}{3} c_L \qquad {\rm and} \qquad \frac{R}{G} = \frac{2}{3} c_R.
\end{equation}
Here $G$ is the bulk Newton's constant and $L , R$ are the AdS radii on the two sides.

We will adopt the familiar Poincar\'{e} patch coordinates $(x,z)$ on both sides of the brane:
\begin{equation}
ds^2 = L^2 \frac{-dt^2 + dx^2 + dz^2}{z^2}
\label{eq:poincare}
\qquad {\rm and} \qquad
ds^2 = R^2 \frac{-dt^2 + dx^2 + dz^2}{z^2}
\end{equation}
The wall occupies a surface of constant extrinsic curvature, which in this coordinate system turns out to be a `straight line' in the $z$-$x$ plane. Each patch of AdS$_3$ on one side of the wall is characterized by the slope of that line, which we express in terms of $\alpha$ and $\beta$:
\begin{equation}
z = -x \tan\beta~{\rm (left)}
\qquad {\rm and} \qquad
z = x \tan\alpha~{\rm (right)}
\end{equation}
Fig.~\ref{fig:gluing} depicts one example geometry, in which $\alpha$ and $\beta$ are both less than $\pi/2$. Note that $\alpha = \pi/2$ denotes one half of the hyperbolic disk delimited by a radial geodesic. Thus, the na{\"\i}ve holographic interpretation of minimal updates predicts that $\alpha, \beta \geq \pi/2$.

We now verify that the thin wall models conform to this prediction.

In the thin wall geometry, Einstein's equations reduce to the Israel junction conditions \cite{israel}, which we re-derive in Appendix~\ref{app:thinwallaction}. For a brane of tension $\lambda$, these take the form:
\begin{equation}
\frac{L}{\sin \beta} = \frac{R}{\sin\alpha} = -\frac{\cot\alpha + \cot\beta}{8\pi G\lambda}\,.
\label{eq:israel}
\end{equation}
These three quantities are equal to the radius of intrinsic curvature on the brane. Observe that eqs.~(\ref{eq:israel}) accommodate the duals of boundary CFTs discussed in \cite{Takayanagi:2011zk, Fujita:2011fp, ugajin} simply by setting $\beta = \pi/2$. This introduces a fictitious left chunk of AdS$_3$ with curvature $L = R/\sin\alpha$ which decouples, because it exerts no force on the bulk wall.

Although eqs.~(\ref{eq:israel}) have formal solutions with arbitrary $\alpha$ and $\beta$, in fact only $\alpha, \beta \geq \pi/2$ are physical. When $\alpha, \beta < \pi/2$, the tension $\lambda$ is forced to be negative, which violates the weak energy condition in the bulk.\footnote{Ref.~\cite{erdmenger} contains a thorough discussion of energy conditions in the context of holographic dCFTs.} Such a situation gives rise to rather exotic features associated with strong subadditivity, which we detail in Appendix~\ref{snelletc}. 

The remaining case, $\alpha \geq \pi/2 > \beta$, is also unphysical. As we show in Appendix~\ref{app:thinwallaction}, in this regime the wall is necessarily unstable so it cannot be the dual of the ground state of a dCFT.

Studying geodesics in the thin wall space-time built by a wall with positive tension turns out to involve an interesting application of Snell's law. Because we have not found a solution of this problem anywhere in the literature, in Appendix~\ref{snelletc} we explain how to find such geodesics and compute the kinematic space of the thin wall geometry.

\paragraph{Summary} The thin wall geometry is consistent with the na{\"\i}ve holographic interpretation of the minimally updated MERA. This is true regardless of whether we adopt the direct \cite{swingle} or the kinematic \cite{secondpaper} proposal for relating MERA to holographic geometries.

However, the direct proposal is arguably subject to some awkward caveats. This is because $\alpha, \beta > \pi/2$ means that the thin wall geometry is strictly larger than it would have been in the absence of a defect. Thus, the causal cone of the defect must be simultaneously interpretable as the radial geodesic (in the dual of the undeformed CFT) and as the extra thickness of space-time grown by the thin wall (quantified by $\alpha + \beta -\pi$.) This caveat does not arise in the kinematic proposal where, with or without the wall, we are always dealing with the same set of geodesics. We will not dwell on this issue further because more general models will anyway force us to revise our assumptions.

\subsection{Thick Walls: Not All Bonds Are Created Equal}
\label{weighingbonds}

The exercise of studying thin wall models is useful because it immediately illustrates why the `na{\"\i}ve holographic interpretation' of the minimally updated MERA is na{\"\i}ve. As soon as our wall is no longer thin, it will involve non-trivial profiles of various bulk fields whose tails extend all the way to the asymptotic boundary. Indeed, the non-vanishing one-point functions~(\ref{eq:1pt}) of holographic dCFTs are read off precisely from such tails of normalizable modes of bulk fields. Looking for two greater-than-half chunks of pure AdS$_3$ on both sides of the wall can only work in a thin wall model.

There is another reason why the na{\"\i}ve interpretation is too na{\"\i}ve. When we discussed the direct \cite{swingle} and the kinematic \cite{secondpaper} readings of MERA, the full $SO(2,2)$ symmetry of the theory appeared to be a key ingredient. In the direct proposal, the connection between minimal cuts in MERA and geodesics in AdS$_3$ was only sensible because every MERA bond contributed an equal amount to the entropy count \cite{mera}. This feature relies on the global $SO(2,2)$ symmetry. To see this, recall that changing the UV cut in MERA corresponds to applying a conformal transformation \cite{BTZpaper}. Any bond in MERA can become a part of the UV cut under the action of $SO(2,2)$ and therefore all bonds are related to one another by this symmetry. In the kinematic proposal, on the other hand, the $SO(2,2)$ entered via the choice of measure $\mathcal{D}{\rm (geodesics)}$, which translated into uniformly counting different MERA tensors.

In the case at hand, the symmetry is broken to $SO(2,1)$. On the spatial slice modeled by the tensor network, the only symmetry we have are dilations about the origin. In order to relate thick wall models to MUP, we must assign different weights to different tensors and bonds in the minimally updated MERA.

\paragraph{Assigning relative weights to bonds and tensors} Ref.~\cite{secondpaper} explained how to weigh different regions of MERA in the kinematic interpretation. To explain this prescription, we need a few basic facts.

In the present context, the kinematic space is the space of intervals on a spatial slice of a CFT$_2$. When a holographic dual is available, it is also the space of bulk geodesics. The kinematic space has a Lorentzian metric of the form:
\begin{equation}
ds^2_{\rm K.S.} = \frac{\partial^2 S_{\rm ent}(u,v)}{\partial u \partial v}\, dudv\,,
\label{ksmetric}
\end{equation}
where $u$ and $v$ are the two endpoints of a CFT interval / bulk geodesic and $S_{\rm ent}$ is the entanglement entropy of the interval / length of the geodesic. This metric turns out to be de Sitter space in the case of a locally AdS geometry, and has many attractive properties which were discussed in \cite{kinematicsp, secondpaper} and elsewhere \cite{stereoscopy, deBoer:2016pqk,Asplund:2016koz, chenbin}. For example, the volume form derived from this metric defines a measure on the space of bulk geodesics $\mathcal{D}{\rm (geodesics)}$ such that eq.~(\ref{eq:crofton}) holds.

The claim of \cite{secondpaper} is that we can think of MERA as a discrete version of kinematic space. To do so, consider two pairs of nearby points, $(u, u-\Delta u)$ and $(v, v+\Delta v)$, on the UV cut of MERA. We can impose on MERA a discretized version of metric~(\ref{ksmetric}):
\begin{equation}
ds^2_{\rm MERA} =
S_{\rm ent}(u - \Delta u, v) + S_{\rm ent}(u, v+\Delta v) - S(u-\Delta u, v + \Delta v) - S(u,v)\,.
\label{merametric}
\end{equation}
In this `metric', the light-like directions $u$ and $v$ agree with the causal structure of MERA, which we mentioned in Sec.~\ref{sec:minupdates}. The quantity (\ref{merametric}) coincides with a discretized `volume form' on the tensors of MERA, which can be compared with $\mathcal{D}{\rm (geodesics)}$.

In the ground state of an $SO(2,2)$-invariant theory, eq.~(\ref{merametric}) defines a discrete version of two-dimensional de Sitter space.\footnote{For other observations relating MERA to de Sitter space, see \cite{Beny:2011vh, SinaiKunkolienkar:2016lgg}.} But in a theory with only $SO(2,1)$ invariance, the `volumes' assigned to different regions of MERA will differ. The only fact guaranteed by the symmetry is that identical regions living on the same ray (as discussed in Sec.~\ref{sec:rayedMERA}) carry equal volumes.

\paragraph{A case study in thick walls: the AdS$_3$-Janus solution} One holographic pair which illustrates this non-uniformity is the Janus deformation of AdS$_3$ and its dual interface CFT. Following earlier developments in AdS$_5$ \cite{Bak:2003jk}, Refs.~\cite{Bak:2007jm, Azeyanagi:2007qj, Chiodaroli:2010ur, Bak:2011ga, gutperle2015, bakrecent} studied a scalar field $\phi$ (the `dilaton') coupled to Einstein gravity with a negative cosmological constant in three dimensions and found the following solution:
\begin{align}
ds^{2} & = L^2 \left( du^{2} + \rho(u)^2\, ds_{AdS_{2}}^{2} \right)
\label{janusgen}
\\
ds_{AdS_{2}}^{2} & = -\cosh^{2}{r}dt^{2} + dr^{2} \\
\rho(u)^2 & = \frac{1}{2}(1+\sqrt{1-2\gamma^{2}}\cosh{2u})
\label{janusrho} \\
\phi{(u)} & = \phi_{0} + \frac{1}{\sqrt{2}}\log{\left(\frac{1+\sqrt{1-2\gamma^{2}} + \sqrt{2}\gamma\tanh{u}} {1+\sqrt{1-2\gamma^{2}} - \sqrt{2}\gamma\tanh{u}}\right)}
\label{janusphi}
\end{align}
They also explained how this solution is holographically dual to the ground state of a marginal deformation of the D1-D5 CFT whose strength is proportional to $\gamma$. The deformation has a different sign on the two halves of the boundary, so the resulting theory is an interface CFT.\@ In the bulk, the AdS$_3$-Janus solution contains a thick wall.

We do not have an optimized tensor network which prepares the ground state of this theory, so we cannot make quantitative comparisons with MERA.\@ But we can compute its kinematic space (eq.~\ref{ksmetric}) and observe qualitative features. We carried out this computation for small $\gamma$ in Appendix~\ref{App:Janus}. Up to an overall factor of $L/2G$, the result to first non-trivial order in $\gamma$ reads:
\begin{equation}
ds^2_{\rm K.S.-Janus} =
\frac{du\, dv}{(u-v)^2} \left[ 1 -\frac{\gamma^{2}}{2}
\bigg(
\eta^2 + 3 - \frac{1}{2}
\left(\eta^3 + 3\eta^{-1} \right)\log{\left|\frac{1+\eta}{1-\eta}\right|} \bigg)\right]
\label{eq:janusks}
\end{equation}
Here $\eta = (v-u)/(v+u)$ is a kinematic $SO(2,1)$ invariant, related to $\xi$ from eq.~(\ref{eq:defxi}) via:
\begin{equation}
\eta = (\xi^{-1} + 1)^{-1}\,.
\end{equation}

The inside of the causal cone of the interface has $\eta > 1$ while the regions in MERA that are imported from the parent without updates have $\eta < 1$. Indeed, the effect of the interface spills out beyond the causal cone of the interface, and increases the kinematic volume there. It is UV-finite and in fact vanishes in the UV limit $\eta \to 0$, where the effect of the interface is the smallest.

Within the causal cone, on the other hand, the interface causes the overall kinematic volume to decrease. This is to be expected because according to eq.~(\ref{eq:crofton}) the volume of this region computes the entanglement entropy of the two sides of the interface.

\paragraph{Summary:} The bulk duals of holographic dCFTs generically involve thick walls. In relating such theories to tensor networks, we cannot count all tensors or bonds with equal weight. Instead, we must account for different weights that occur at different values of the $SO(2,1)$ invariant $\xi$ (see eq.~\ref{eq:defxi}). In a minimally updated MERA, even though all tensors outside the causal cone are identical, their weights differ depending on the location relative to the defect.

\subsection{Non-normalizable Modes: From the Minimally Updated MERA to Rayed MERA}

The above conclusion poses one residual question. On the one hand, the MUP mandates that some tensors do not register the presence of a defect; on the other hand, those tensors count with different weights when we calculate entropies. What then distinguishes states constructible using the minimally updated MERA versus the rayed MERA? We would like to answer this question in a way that makes contact with the AdS/CFT correspondence.

Recall that the minimally updated MERA is designed for theories constructed by coupling two $SO(2,2)$ invariant parent theories along a common interface. The rayed MERA is for a generic $SO(2,1)$-invariant theory, which could be constructed in multiple ways. One such way is to deform a parent theory by an appropriately selected source, which is either $SO(2,1)$-invariant or designed to recover the $SO(2,1)$ after an RG flow. In holography, deforming theories by the introduction of sources is effected by turning on non-normalizable modes in the bulk \cite{nonnorm}. Thus, a ground state of a holographic theory whose bulk dual involves a thick wall can be prepared by either one of the two types of networks---the minimally updated MERA or the rayed MERA---depending on whether the thick wall contains condensates of non-normalizable modes away from the `interface.' Here by `interface' we mean the fixed world-line of the residual $SO(2,1)$ symmetry.

As an example, the holographic dual of the AdS$_3$-Janus solution is a marginal deformation of the D1-D5 CFT \cite{Bak:2007jm}:
\begin{equation}
S = S_{\,{\rm D1D5}}
+ \tilde{\gamma} \int_{x>0}\!\!dx\,dt\,\mathcal{O}_\phi(x,t)
- \tilde{\gamma} \int_{x<0}\!\!dx\,dt\,\mathcal{O}_\phi(x,t)
\label{janusdeform}
\end{equation}
Here $\tilde{\gamma}$ is a deformation parameter, which agrees with the $\gamma$ from eqs.~(\ref{janusrho}) and (\ref{janusphi}) to leading order, $\tilde{\gamma} = \gamma + O(\gamma^2)$. The bulk solution involves a non-normalizable mode for the dilaton, which asymptotes to different constant values on the boundary
\begin{equation}
\phi \to \phi_\pm = \phi_0 \pm \frac{1}{\sqrt{2}}\tanh^{-1} \sqrt{2}\gamma
\end{equation}
and accounts for the deformation (\ref{janusdeform}).

Eq.~(\ref{janusdeform}) is a marginal deformation of the parent CFT with a piece-wise constant source that jumps at the interface. If, in principle, we had at our disposal MERA representations of the ground states of the theories
\begin{equation}
S = S_{\,{\rm D1D5}}
\pm \tilde{\gamma} \int_{{\rm all}~x}\!\!dx\,dt\,\mathcal{O}_\phi(x,t)\,,
\end{equation}
we could use them as input in the minimal updates prescription. Thus, the ground state of the theory dual to the AdS$_3$-Janus solution belongs to the class of states, which can in principle be represented in the form of a minimally updated MERA.\@ Of course, the tensors comprising that network would be different from those which prepare the ground state of the undeformed theory.

However, if we turn on more general deformations while preserving $SO(2,1)$, the resulting ground states can only be prepared using the rayed MERA. For example, we could deform a holographic CFT with irrelevant operators coupled to sources with a power-law dependence on the distance from a select line. If the interior of the resulting bulk geometry were then compared to a MERA-type tensor network, it would have to be a rayed MERA.

\paragraph{Summary:} The distinction between the minimally updated MERA and the rayed MERA is whether we simply couple two parent CFTs along an interface or do something more generic. One option in the latter category is to deform the theory globally by an irrelevant operator to find a new fixed point in the IR.\@ Such a theory is generally outside the scope of the minimal updates prescription, but if it preserves $SO(2,1)$, it can in principle be captured by a rayed MERA.\@

\section{Discussion}
\label{sec:discussion}

There is now a considerable literature which seeks ways to relate spacetimes that arise in holographic duality to tensor networks. In this paper, we initiate a new chapter of this endeavor: studying space-times which are neither pure anti-de Sitter nor its quotients nor Virasoro descendants. For this initial study we chose to consider holographic defect, interface and boundary CFTs (dCFTs) and tensor networks in the class of the Multi-scale Entanglement Renormalization Ansatz (MERA).

We concentrated on MERA for 1+1-dimensional CFTs because this class of networks is best understood. In particular, in MERA we know (a) how to realize conformal transformations (by changing the UV cut \cite{BTZpaper}), (b) how the spectrum of conformal dimensions and OPE coefficients are encoded (for details, see \cite{meraope}), and (c) how to represent ground states of dCFTs (the minimal updates proposal \cite{minupdates}). Concerning the class of theories, we focused on dCFTs because they obey a residual $SO(2,1)$ global symmetry, which has a clarifying power. It organizes data in both MERA (on rays emanating from the origin) and in the holographic geometry (which is foliated by AdS$_2$ slices.)

Some of our conclusions concern specifically the MERA class of tensor networks. We clarified and complemented arguments supporting the validity of the minimal updates proposal (Sec.~\ref{sec:minupdates}) and proposed an extension for generic, $SO(2,1)$-invariant theories (rayed MERA, Sec.~\ref{sec:rayedMERA}). Our other conclusions should hold more generically. In particular, we expect that in every meaningful instance of a holographic bulk geometry-tensor network correspondence, the following rule should hold:
\begin{itemize}
\item Changing tensors in the ground state network represents turning on non-normalizable modes in the bulk. 
%In AdS/CFT, turning on non-normalizable modes deforms the CFT, which in general changes its spectrum and OPE coefficients. Because this CFT data should be encoded in the ground state network, altering its tensors should cleanly represent the impact of non-normalizable modes.
\end{itemize}
In the case of MERA, because of its causal structure, the effect of locally turning on a non-normalizable mode is contained in the causal cone of the deformation. We propose this as the holographic interpretation of the theory of minimal updates \cite{minupdates}. But in other types of networks such as \cite{Pastawski:2015qua,Yang:2015uoa,randomtensors, janet, Donnelly:2016qqt}, the effect of a deformation should also be cleanly identifiable and likely localized in a subregion of the network.

%Holographically, we can think of this rule in the following way. The bulk low energy effective field theory (LEEFT) implicitly encodes the OPE coefficients of the dual CFT.\@ When the CFT is deformed, the bulk registers the deformation through changes in its LEEFT.\@ Anticipating that tensors in the ground state network change when a deformation is applied reflects the intuition that the tensors encode bulk physics.

At the same time, we should remember that local properties of a tensor network state in general depend non-locally on the tensors. One example considered in this paper (see Sec.~\ref{weighingbonds}) is the set of entanglement entropies, which underlie both the direct~\cite{swingle} and the kinematic~\cite{secondpaper} holographic interpretation of MERA.\@ We can think of such local but non-locally determined properties of tensor network states as akin to the normalizable bulk modes. In AdS/CFT, these encode responses to boundary conditions set elsewhere. Other familiar examples of such quantities are CFT one-point functions, which in MERA depend on the entire causal future of the given point.

\paragraph{Next steps} It would be interesting to realize some of these ideas in other types of tensor networks, which were specifically designed for the AdS/CFT correspondence~\cite{Pastawski:2015qua,Yang:2015uoa,randomtensors, janet, Donnelly:2016qqt}, and also consider the Kondo problem as an example \cite{Erdmenger:2015spo}. Many questions await answers: How do these networks encode OPE coefficients of the CFT?\@ Can we see how deforming the CFT changes the ground state tensors and thus observe the effect of a non-normalizable mode? How to represent ground states of defect CFTs? More specifically, how to deform those networks to construct an analogue of a thin wall geometry? This last problem is further pertinent for understanding how those classes of tensor networks can accommodate the backreaction of bulk matter fields.

Departing from tensor networks, our paper is the first study of the kinematic space of dCFTs. For ordinary CFTs, studying fields \emph{local in kinematic space} led to enlarging the holographic dictionary by the addition of OPE blocks, which at leading order in $1/N$ are dual to bulk fields integrated along geodesics \cite{stereoscopy, deBoer:2016pqk}. It would be interesting to generalize these findings to holographic dCFTs, perhaps starting with thin wall bulk duals. Interesting work in this direction is forthcoming \cite{jamieinprogress}.

\section*{Acknowledgments}
We thank Xi Dong, Glen Evenbly, Markus Hauru, Micha{\l} Heller, Lampros Lamprou, Samuel McCandlish, Ashley Milsted, Rob Myers, James Sully and Guifr{\'e} Vidal for helpful discussions, comments and insights.
BC is supported by the Peter Svennilson Membership in the Institute for Advanced Study while PN and SS are supported by the National Science Foundation under Grant Number PHY-1620610. SS would like to thank the Perimeter Institute for Theoretical Physics (PI) for hospitality during part of this work. Research at PI is supported by the Government of Canada through Innovation, Science and Economic Development Canada and by the Province of Ontario through the Ministry of Research, Innovation and Science. BC dedicates this paper to Bayu Mi{\l}osz Czech and his patient mother, Stella Christie.

\appendix

\section{Israel Junction Conditions and Wall Stability}
\label{app:thinwallaction}

We consider three-dimensional geometries, which preserve $SO(1, 2)$ symmetry:
\begin{equation}
ds^2 = du^2 + \rho(u)^2 (-\cosh^2 r dt^2 + dr^2)
\label{template}
\end{equation}
For a dual of a general holographic dCFT, we should also include other fields and their backreactions; one such example is discussed in Sec.~\ref{weighingbonds} and Appendix~\ref{App:Janus}. Here we assume that the geometry contains a thin wall of tension $\lambda$. To have a locally AdS$_3$ geometry to the left of the wall, we must have
\begin{equation}
\rho(u) = L \cosh (u/L),
\label{formrho}
\end{equation}
where $L$ is the left AdS$_3$ curvature radius. To the right of the wall, we will have a similar expression with $L \to R$, the curvature radius on the right. On the static slice $t = 0$, the change of coordinates from (\ref{template}) to (\ref{eq:poincare}) is:
\begin{equation}
z = e^{r}{\rm sech}\,{u/L} \qquad {\rm and} \qquad
x = -e^{r}\tanh{u/L}\,.
\label{coordchange}
\end{equation}
Away from a spatial slice the formulas are more involved, but we do not need them in this paper.

In eq.~(\ref{formrho}), the asymptotic boundary of space-time is approached as $u \to -\infty$ while the wall sits at some specific value $u_*$. The $u=0$ slice of metric (\ref{template}) is a minimal surface in AdS$_3$, so depending on the sign of $u$ the constant-$u$ slices are contracting (for $u<0$) or expanding (for $u>0$) in the direction of increasing $u$, that is toward the wall. This distinction will be important for our considerations.

To find a static configuration of the AdS$_{3}$ chunks and the wall, we consider the Einstein-Hilbert action with a Gibbons-Hawking-York (GHY) term and an explicit wall contribution:
\begin{equation}
S = \frac{1}{16\pi G} \int_{\rm left} d^3x \sqrt{-g} (\mathbf{R} - 2 \Lambda) + \frac{1}{8\pi G} \int_{\rm wall} d^2y \sqrt{-h} K_L + (L \to R) -\lambda \int_{\rm wall} d^2y \sqrt{-h}
\label{action3}
\end{equation}
Additional GHY terms arise at the asymptotic boundary of space-time, but these will play no role in our analysis. The Ricci scalar in metric (\ref{template}) takes the form:
\begin{equation}
\mathbf{R} = -2 \, \frac{1 + \rho'^2 + 2 \rho \rho''}{L^2 \rho^2}
\label{ricciscalar}
\end{equation}
We can confirm the correctness of this expression by substituting (\ref{formrho}), which gives $\mathbf{R} = -6/L^2$.
Plugging eq.~(\ref{ricciscalar}) and $\Lambda = -L^{-2}$ into (\ref{action3}), the action takes the form:
\begin{equation}
S \propto -\frac{L}{8\pi G} \int^{u_*} du \left(1 + \rho'^2 + 2 \rho \rho''- \rho^2\right) + \frac{L^2 {\rho(u_*)}^2 K_L}{8\pi G} + (L \to R) - L^2 {\rho(u_*)}^2 \lambda
\label{action2}
\end{equation}
Here we have dropped an overall infinite factor, which stands for the volume of AdS$_{2}$ with unit curvature.

Expression (\ref{action2}) contains two terms, which can be combined and simplified. To get a standard variational problem, we need to eliminate $\rho''$ via integration by parts. This introduces a boundary term, which the GHY term is designed to cancel:
\begin{equation}
- L \int^{u_*} du\, 2\rho \rho'' + L^2 {\rho(u_*)}^2 K_L =
- L \int^{u_*} du\, 2\rho \rho'' + L \frac{d}{du} \rho^2\Big|_{u_*} =
L \int^{u_*} du\, 2\rho'^2
\end{equation}
After this substitution, action (\ref{action2}) becomes:
\begin{equation}
S =
\frac{L}{8\pi G} \int^{u_*} du \left(\rho'^2 -1 + \rho^2\right) + (L \to R) - L^2 {\rho(u_*)}^2 \lambda
\end{equation}
We may now plug in the known solution (\ref{formrho}) for $\rho(u)$ and its right counterpart to obtain:
\begin{equation}
S = \frac{L}{4\pi G} \int^{u_*} du\, \sinh^2 u
+ \frac{R}{4\pi G} \int^{v_*} dv\, \sinh^2 v
- \lambda\, L^2 \cosh^2 u_*.
\label{action}
\end{equation}

For continuity of the metric, the intrinsic geometry of the wall must be the same in both the $u$ and $v$ metrics. This leads to the first Israel junction condition \cite{israel}, which is the first equality in eq.~(\ref{eq:israel}):
\begin{equation}
L \cosh u_* = R \cosh v_*.
\label{glue}
\end{equation}
Note that we have two distinct branches of $v_*$, which correspond to having a `smaller-than-half' or `bigger-than-half' chunks of AdS$_{3}$ to the right of the wall:
\begin{equation}
\sinh v_* = \pm \sqrt{{(L/R)}^2 \cosh^2 u_* - 1}
\label{vstarustar}
\end{equation}
To the left of the wall, the analogous distinction is controlled by the sign of $u_*$.

It is now trivial to find the equilibrium configuration of the AdS$_{3}$ patches and the wall. Setting $dS/du_* = 0$ gives:
\begin{equation}
\frac{L}{4\pi G} \sinh^2u_* + \frac{R}{4\pi G} \sinh^2 v_* \cdot \frac{dv_*}{du_*} - 2 \lambda L^2 \cosh u_* \sinh u_* = 0
\end{equation}
Substituting
\begin{equation}
\frac{dv_*}{du_*} =
\frac{L \sinh u_*}{R \sinh v_*}
\end{equation}
which follows from (\ref{glue}), we get:
\begin{equation}
\sinh u_* \big( \sinh u_* + \sinh v_* - 8 \pi G \lambda L \cosh u_* \big) = 0.
\label{extrema}
\end{equation}
Setting $u_* = 0$ is not a solution of the equations of motion; rather, it signals a breakdown of $u_*$ as a collective coordinate. Equating the other factor of (\ref{extrema}) to zero gives the second Israel junction condition, which is the second equality in (\ref{eq:israel}).

To check the stability of the solution, we compute:
\begin{equation}
\frac{d^2S}{du_*^2}\,\Bigg|_{\rm EOM} = 2 \lambda L^2\, \frac{\sinh u_*}{\sinh v_*}\,.
\end{equation}
Thus, stability requires that the product of $\lambda$, $u_*$ and $v_*$ must be positive. Excluding negative tensions leaves out $u_*, v_* < 0$ ($\lambda > 0$ forbids this by the equation of motion) and $u_*, v_* > 0$, i.e. $\alpha, \beta > \pi/2$. This is the only consistent, stable configuration.

\section{Geodesics in the Thin Wall Geometry}
\label{snelletc}

It is interesting to find the geodesics of the thin wall geometry explicitly. We denote the endpoints of the geodesic with $a, b$ and assume $a > b$.

\paragraph{Geodesics in the presence of a stable thin wall}
The stable configuration has $\alpha, \beta > \pi/2$. Geodesics that begin and end on the same side of the wall are same as in pure AdS$_3$. Their lengths are
\begin{equation}
S(a,b) = 2L \log\frac{a-b}{\mu}\,,
\end{equation}
where $\mu$ is a large scale cutoff in the geometry. In the following we will drop the cutoffs, which in three bulk dimensions are simple additive constants.

\begin{figure}[!htbp]
    \begin{center}
        \begin{tabular}{cc}
        \includegraphics[width=6cm]{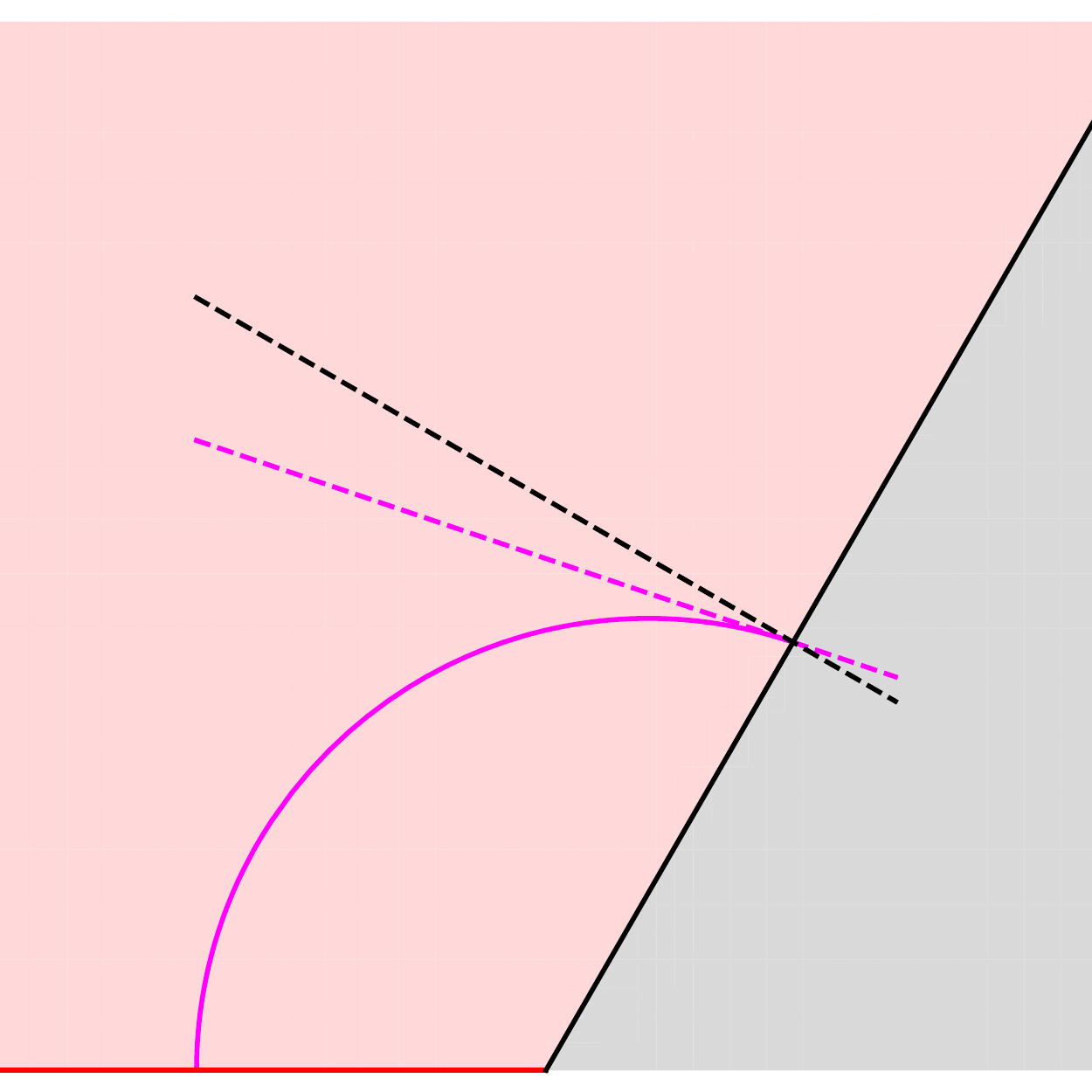} & \includegraphics[width=6cm]{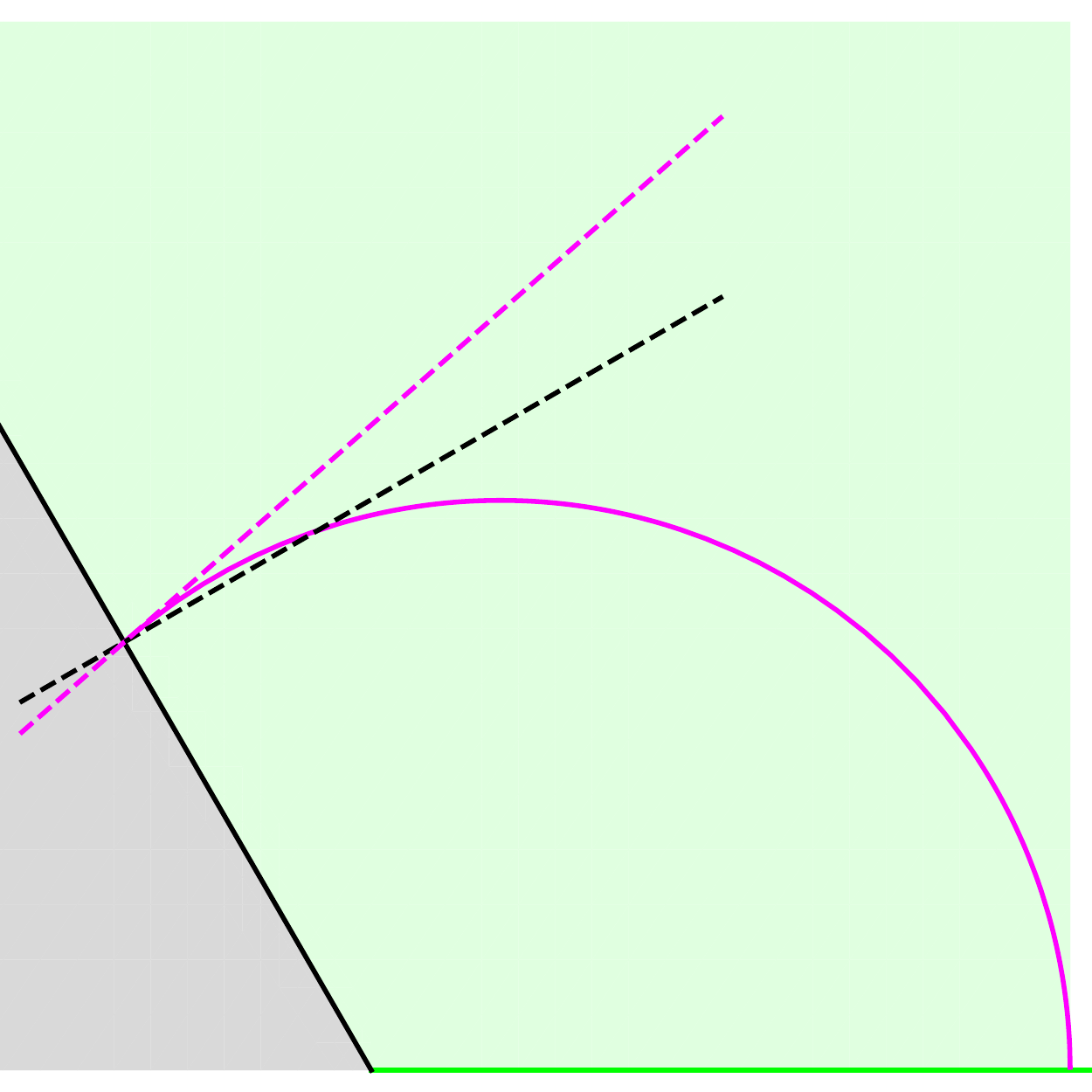}
        \end{tabular}
    \end{center}
	\caption{A wall-crossing geodesic in a thin wall geometry consists of two arcs, which meet the wall at the same angle and location.}
\label{fig:SingleCrossingSymmetric}
\end{figure}

To find the geodesics crossing the wall ($b < 0 < a$), observe that the geodesic motion in the hyperbolic plane is analogous to the propagation of a light ray in a medium whose index of refraction is $n(z) = L/z$.
Due to the first Israel junction condition, the index of refraction at the brane is continuous. Thus, by Snell's law, a geodesic crossing the brane consists of two circular arcs, which meet at the location on the brane where no refraction occurs. The angles can be read off directly from the $x$-$z$ plane, which is conformal to the geometry. Thus, we are looking for two arcs which meet the wall at the same location and the same angle in the $x$-$z$ plane. One such a geodesic is plotted in Fig.~\ref{fig:SingleCrossingSymmetric}.

Finding this location is a simple minimization exercise. Consider a family of piecewise geodesic curves, each of which consists of two circular arcs meeting at an arbitary junction on the brane. Let $y=\sqrt{x^2 + z^2}$ be the coordinate distance of the junction from the defect; note that $y$-values on the two sides of the wall agree. One can easily write down the length of such a curve as a function of $y$:
\begin{equation}\label{S}
S{(\alpha,\beta,a,b,y)} =
L\log{\left(\frac{b^{2}+y^{2}-2|b|y\cos{\beta}}{y\sin{\beta}}\right)}
+
R\log{\left(\frac{a^{2}+y^{2}-2ay\cos{\alpha}}{y\sin{\alpha}}\right)}
\end{equation}
To find the actual geodesic among this family of curves, we minimize the length formula above with respect to $y$. The critical value of $y$, which we denote $y_{*}$, is given by:
\begin{equation}
y_{*} = \frac{1}{2} \csc \left(\frac{\alpha+\beta}{2}\right) \left[ (a-|b|)\sin{\left(\frac{\beta-\alpha}{2}\right)} + \sqrt{(a+|b|)^{2}\sin^{2}{\left( \frac{\beta-\alpha}{2}\right)} + 4a|b|\sin{\alpha}\sin{\beta} } \right].
\end{equation}
Substituting this expression in (\ref{S}) gives the desired geodesic length. For the kinematic space metric component, we would then take the second partial with respect to $a$ and $b$ as in eq.~(\ref{ksmetric}). We do not give the full expression here because it is not illuminating.

\begin{figure}[t]
\begin{center}
    \includegraphics[width=12cm]{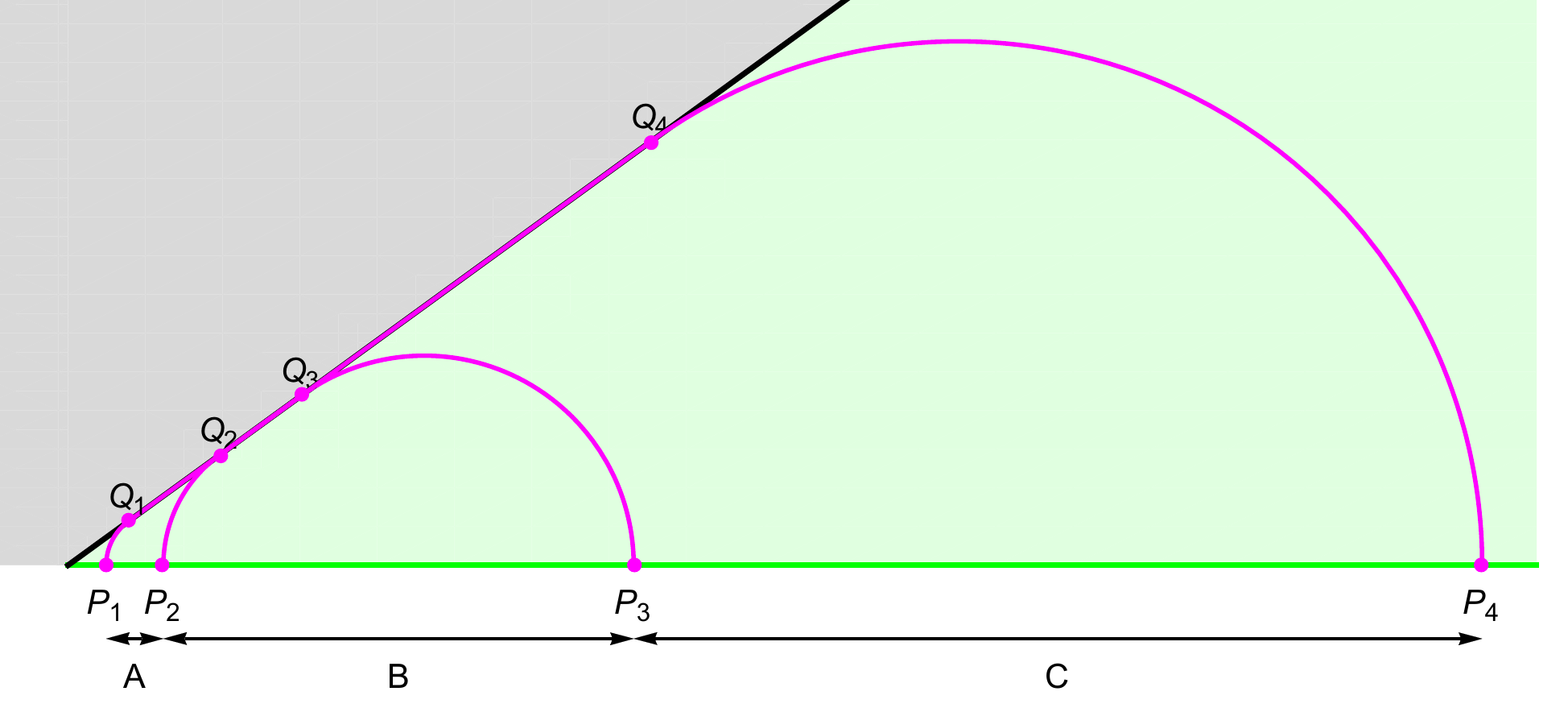}
\end{center}
\caption{Illustration of SSA saturation for squashed geodesics. A family of squashed geodesics spanning three adjacent boundary intervals.}
\label{fig:SSAfig}
\end{figure}

\paragraph{Negative wall tension and strong subadditivity}
The pathological case when both $\alpha , \beta < \pi/2$ has some further exotic properties. Geodesics corresponding to regions with $\xi$ greater than a certain critical value are `squashed' by the wall: they consist of two semi-circular arcs that are tangent to the wall plus a finite segment along the wall. 
A family of such geodesics spanning three adjacent boundary intervals are depicted in Fig.~\ref{fig:SSAfig}.

%\begin{figure}[!h]

If we assume that this geometry obeys the Ryu-Takayanagi proposal for some dual CFT state, we immediately see that intervals depicted in Fig.~\ref{fig:SSAfig} saturate the strong subadditivity (SSA) of entanglement entropy. In kinematic space, SSA saturation results in a degenerate metric in certain wedge-shaped regions near the edges of the defect's  causal cone. Saturation of SSA places a strong constraint on the entanglement structure of a quantum state~\cite{ssasaturate}. Saturating it over a continuous family of intervals in a field theory is a powerful constraint, even if it is subject to $\mathcal{O}(1/N)$ corrections. It would be interesting to prove that such a set-up cannot be realized in a real CFT.

\section{Kinematic Space of the Janus Solution}
\label{App:Janus}
In this appendix, we compute the entanglement entropy and kinematic space of the Janus solution perturbatively for small $\gamma^{2}$. 
%We note that there already exists a large literature on the topic of holographic entanglement entropy in dCFTs; for a selection see \cite{gutperle2015, Azeyanagi:2007qj, Jensen:2013lxa, Estes:2014hka, Sakai:2008tt, Chiodaroli:2010ur, Bianchi:2015liz, Chang:2013mca}. 
On a constant time slice of the Janus solution, expanding the metric~(\ref{janusgen}-\ref{janusrho}) to first non-trivial order in $\gamma$ gives:
\begin{equation}\label{Janussmallgamma}
ds^{2} = L^2 \left(\cosh^{2}{u} - \frac{\gamma^2}{2}\cosh{2u} \right) dr^{2} + L^2 du^{2}
\end{equation}
Applying the coordinate change (\ref{coordchange}) brings this metric to the form:
\begin{equation}\label{Janussmallgamma}
ds^{2} = L^2 \frac{dx^{2}+dz^{2}}{z^{2}} - \gamma^{2}L^2\frac{(z^{2}+2x^{2})}{2z^{2}(x^{2}+z^{2})^{2}}(xdx+zdz)^{2}
\end{equation}
Perturbations of geodesic lengths generally arise from two effects: the shift in the metric and the shift in the coordinate trajectory of the geodesic. To lowest order, however, we can ignore the latter and only consider the former. Thus, we will take the geodesics to be semi-circles in the $x$-$z$ plane. The perturbed induced metric on the semi-circle, which connects $u = x_0 - R$ and $v = x_0 + R$ takes the form:
\begin{equation}
ds^{2} = L^2 \left[\frac{R^{2}}{(R^{2}-(x-x_{0})^{2})^{2}} - \gamma^{2} \frac{x_{0}^{2}(R^{2}+x^{2}+2xx_{0}-x_{0}^{2})}{2(R^{2}+2xx_{0}-x_{0}^{2})^{2}(R^{2}-x^{2}+2xx_{0}-x_{0}^{2})}\right] dx^{2}
\end{equation}
The perturbation of the length is:
\begin{equation}
\delta S = \frac{1}{2} \int \sqrt{g_{xx}} g^{xx} \delta g_{xx} dx
\end{equation}
Evaluating the integral gives:
\begin{equation}
\delta S{(R,x_{0})} = -\gamma^{2}\frac{L x_{0}^{2}}{2R} \int_{x_{0}-R}^{x_{0}+R} \!\!
\frac{R^{2}+x^{2}+2xx_{0}-x_{0}^{2}}{(R^{2}+2xx_{0}-x_{0}^{2})^{2}}dx =
-\frac{\gamma^{2}L}{8Rx_{0}} \left(4Rx_{0} + 2(R^{2}-3x_{0}^{2})\log{\bigg|\frac{R-x_{0}}{R+x_{0}}\bigg|} \right)
\end{equation}
The correction to the kinematic space metric due to $\gamma$ is then found by differentiation of $\delta S$:
\begin{eqnarray}
\frac{\partial^{2}}{\partial u\partial v}\delta S &=& \frac{1}{4}\left(\frac{\partial^{2}}{\partial x_{0}^{2}} - \frac{\partial^{2}}{\partial R^{2}}\right)\delta S \nonumber \\
&=& -\frac{\gamma^{2}L}{16 R^{3}x_{0}^{3}} \left(4Rx_{0}(R^{2}+3x_{0}^{2}) + 2(R^{4}+3x_{0}^{4})\log{\bigg|\frac{R-x_{0}}{R+x_{0}}\bigg|} \right)
\label{januscorr}
\end{eqnarray}
This is eq.~(\ref{eq:janusks}) from the main text.

\end{document}